\documentclass[a4paper,fleqn,usenatbib]{mnras}

\usepackage{newtxtext,newtxmath}

\usepackage[T1]{fontenc}
\usepackage{ae,aecompl}

\usepackage{graphicx}	
\usepackage{amsmath}	
\usepackage{hyperref}

\def\hyphenateAndTtWholeString #1{\xHyphenate#1$\wholeString\unskip}

\def\xHyphenate#1#2\wholeString {\if#1$%
    \else\transform{#1}%
    \takeTheRest#2\ofTheString\fi}

\def\takeTheRest#1\ofTheString\fi
{\fi \xHyphenate#1\wholeString}

\def\transform#1{\url{#1}\hskip 0pt plus 1pt}

\def\urlx #1{\href{#1}{\hyphenateAndTtWholeString{#1}}}

\newcommand{\Teff}{T_{\rm eff}\,}

\title[]{Serendipitous discovery of a dusty disc around WDJ181417.84$-$735459.83}

\author[E. Gonz\'alez Egea et al.]{
E. Gonz\'alez Egea$^{1}$\thanks{E-mail: e.gonzalez-egea@herts.ac.uk}, 
R. Raddi$^{2}$, 
D. Koester$^{3}$, 
L. K. Rogers$^{4}$, 
F. Marocco$^{5,6}$,
W. J. Cooper$^{1,7}$,
\newauthor J. C. Beamin$^{8}$,
B. Burningham$^{1}$,
A. Day$-$Jones$^{9}$,
J. Forbrich$^{1}$ 
and D. J. Pinfield$^{1}$
\\
$^{1}$ Centre for Astrophysics Research, University of Hertfordshire, Hatfield, AL10 9AB, United Kingdom\\
$^{2}$ Departament de F\'isica, Universitat Polit\`ecnica de Catalunya, c/Esteve Terrades 5, E-08860 Castelldefels, Spain\\
$^{3}$ Institut für Theoretische Physik und Astrophysik, Christian-Albrechts-Universität, 24118 Kiel, Germany\\
$^{4}$ Institute of Astronomy, University of Cambridge, Madingley Road, Cambridge United Kingdom, CB3 0HA \\
$^{5}$ IPAC, Mail Code 100-22, California Institute of Technology, 1200 E. California Blvd., Pasadena, CA 91125, USA\\
$^{6}$ Jet Propulsion Laboratory, California Institute of Technology, 4800 Oak Grove Drive, M/S 169-327, Pasadena, CA 91109, USA\\
$^{7}$ Istituto Nazionale di Astrofisica, Osservatorio Astrofisico di Torino, Strada Osservatorio 20, I-10025 Pino Torinese, Italy\\
$^{8}$ N\'ucleo de Astroqu\'imica y Astrof\'isica, Instituto de Ciencias Qu\'imicas Aplicadas, Facultad de Ingenier\'ia, Universidad\\ Aut\'onoma de Chile, Av. Pedro de Valdivia 425, 7500912 Santiago, Chile \\
$^{9}$ Earl Mortimer College, South Street, Leominster, Herefordshire, HR6 8JJ, United Kingdom \\
}

\date{Accepted XXX. Received YYY; in original form ZZZ}

\pubyear{2020}

\begin{document}

\label{firstpage}
\pagerange{\pageref{firstpage}--\pageref{lastpage}}
\maketitle

\begin{abstract}
Spectroscopic observations of white dwarfs reveal that many of them are polluted by exoplanetary material, whose bulk composition can be uniquely probed this way. We present a spectroscopic and photometric analysis of the DA white dwarf WDJ181417.84$-$735459.83, an object originally identified to have a strong infrared excess in the 2MASS and \textit{WISE} catalogues that we confirmed to be intrinsic to the white dwarf, and likely corresponding to the emission of a dusty disc around the star. The finding of Ca, Fe and Mg absorption lines in two X-SHOOTER spectra of the white dwarf, taken 8 years apart, is further evidence of accretion from a dusty disc. We do not report variability in the absorption lines between these two spectra. Fitting a blackbody model to the infrared excess gives a temperature of 910$\pm50$ K. We have estimated a total accretion flux from the spectroscopic metal lines of $|\dot{\rm M}| = 1.784 \times 10^{9}\, $g s$^{-1}$. 
\end{abstract}






\section{Introduction} 

White dwarfs, the final evolutionary stage of low to intermediate mass stars, are compact objects whose photospheres typically show absorption spectral lines of hydrogen, helium or do not show any lines (DA, DB or DC white dwarf spectral type, respectively). Due to their high surface gravities, with $\log{g} \sim 8$ [cm s$^{-2}$], elements heavier than H and He rapidly diffuse from the convective envelope into the interior, with timescales much shorter than the white dwarfs' cooling time \citep[e.g.][]{Fontaine1979, Paquette1986, Koester2009}. 

However, traces of metals are found in white dwarfs' spectra as absorption lines, in which case the white dwarfs are classified as DAZ, DBZ or DZ. These metals were initially thought to come from accretion of the interstellar medium \citep[e.g.][]{Bruhweiler1981, Aannestad1985, Dupuis1993}, but were later alternatively explained by ongoing accretion from exoplanetary material around the white dwarf \citep{Farihi2010a}. To date, we know of several hundreds of white dwarfs with atmospheric metals, representing a percentage of 25$-$50 per cent of the spectroscopically confirmed white dwarfs \citep[e.g.][]{Zuckerman2003, Koester2005, Zuckerman2010, Hollands2017}. Dusty discs have been observed as an excess in the infrared (IR) emission of the white dwarf in dozens of these cases, in around 1$-$4 per cent of white dwarfs \citep[e.g.][]{Farihi2009, Debes2011a, Barber2014, Rocchetto2015}. In a few systems, Ca {\sc ii} triplet and other metallic emission lines are also detected in the white dwarf's spectrum, associated with a gaseous component of the accretion disc \citep[e.g.][]{Gansicke2006, Manser2020}. 

 The current widely-accepted explanation for the formation of circumstellar discs around white dwarfs is the tidal disruption of asteroids \citep{Jura2003}. In this model, either a large asteroid or several small ones \citep{Wyatt2014} are tidally disrupted when entering the Roche radius of the white dwarf, forming an opaque flat ring or rings of dust with an extension of less than 1 solar radius, in a Saturn-like ring system.  This \textit{flat ring model} predicts the flux from the disc integrating the Planck function along the ring radius, with a radius-temperature dependence. The model successfully accounts for observations out to 24 microns taken by \textit{Spitzer} space telescope \citep{Fazio2004, Hora2004}, with the caveat that 8 micron photometry can be contaminated in some cases by a silicate emission feature. This scenario includes also the possibility of planetesimals or any other exoplanetary material to be the parent bodies of these discs.  

Earth-sized white dwarfs are intrinsically faint, and their observed discs are compact. It may be that many stars host discs that lie below the detection threshold \citep{Rocchetto2015}. Furthermore, IR excess may also be explained by a substellar companion in the system \citep[e.g.][]{Probst1983}, so a spectrum of the white dwarf is usually needed for the discrimination between the two scenarios, as discs have been found only around metal polluted white dwarfs. 

Variability in the IR, first detected by \citet{Xu2014} and suspected to be present in the majority of white dwarfs with discs \citep{Swan2019, Swan2020} is evidence of dynamical dust evolution, with no generally accepted explanation yet. \citet{Rogers2020} do not find variability in a near infrared monitoring campaign of a sample of white dwarfs with IR excesses, one interpretation being that tidal disruption events are rare and occur on short time-scales. There have been reported changes in spectral lines associated to circumstellar gas \citep[e.g.][]{Manser2016a, Manser2016b, Dennihy2018}. No definitive evidence for variability of absorption lines of metal pollutants have been found \citep{VonHippel2007, Debes2008}.

The analysis of the metal lines and discs found in white dwarfs can give us unique information about compositions of rocky planets \citep[e.g.][]{Jura2014}, asteroids, comets and even gaseous giants (see recent discovery by \citeauthor{Gansicke2019} \citeyear{Gansicke2019}). The dust composition is in some cases consistent with carbon-deficient and rocky material, likely similar to the material of the inner Solar System \citep[e.g.][]{Jura2009a}. We can infer the rate of the accretion of dusty material and abundances of metal species with respect to H or He abundances in white dwarfs' photospheres, whilst the accretion rates are consistent with parent bodies with sizes of the order of kilometers (\citeauthor{ReviewFarihi2016} \citeyear{ReviewFarihi2016} and references therein). The most prevalent elements found in white dwarfs to date are oxygen, magnesium, aluminium, silicon, calcium and iron \citep[e.g.][]{Visscher2013, Gansicke2012, Xu2019}, elements that dominate also in the rocky bodies of the Solar System.

Here we report the discovery of photometric infrared excess and absorption lines due to accretion of metals in a DAZ white dwarf, identified as WDJ181417.84$-$735459.83 by \citet{GentileFusillo20} (hereafter, WDJ1814$-$7354). In Section \ref{sec:discovery} we report the discovery of this object and present photometry and astrometry from the literature. In Section \ref{sec:observations} we present follow-up observations: two epochs of X-SHOOTER spectroscopy in Section \ref{sec:X-Shooter} and \textit{Spitzer} photometry in Section \ref{sec:spitzer}. In Section \ref{sec:analysis} we discuss the analysis of the data: a white dwarf model fit to the spectroscopy in Section \ref{sec:spectroscopicfit}, determination of the composition of the accreted material and the diffusion timescales in Section \ref{sec:abundances} and a disc model fit to the infrared excess in Section \ref{sec:discfit}. We discuss our results in Section \ref{sec:discussion}. The conclusions are presented in Section \ref{sec:conclusions}.


\section{Discovery of WDJ1814$-$7354}
\label{sec:discovery} 

\begin{table} 
\centering
\caption{\textit{Gaia} DR2 astrometry and photometry of the object WDJ1814$-$7354, along with its $\Teff$, $\log{g}$ and distance (d) estimated in section \ref{sec:spectroscopicfit} in parsecs. Barycentric right ascension ($\alpha$) and declination ($\delta$) are in the International Celestial Reference System and at \emph{Gaia} DR2 reference epoch, 2015.5. Proper motion in right ascension ($\mu_{\alpha}\cos{\delta}$) and declination ($\mu_{\delta}$) and parallax ($\varpi$) are also at \emph{Gaia} DR2 reference epoch, 2015.5. Apparent magnitudes are in the three \emph{Gaia} DR2 passbands. }
\label{tab:gaiainfo}
    \begin{tabular}{lc} 
    \hline
    Gaia Source ID  & \textit{Gaia} DR2 6417955993895552128 \\
    \hline
$\alpha$ [deg] & 273.57335015736  \\  
$\delta$ [deg] & $-$73.91738768154  \\
$\mu_{\alpha}\cos{\delta}$ [mas yr$^{-1}$] & $-$63.89$\pm$0.08 \\ 
$\mu_{\delta}$ [mas yr$^{-1}$] & $-$178.2$\pm$0.1 \\ 
$\varpi$ [mas] & 15.51$\pm$0.07 \\ 
G [mag] & 16.2028$\pm$0.0009 \\ 
G$_{\rm BP}$ [mag] &  16.233$\pm$0.005 \\
G$_{\rm RP}$ [mag] &  16.104$\pm$0.004 \\
G$_{\rm BP}$ $-$ G$_{\rm RP}$ [mag] & 0.129$\pm$0.009 \\ 
d [pc] & 64.6$\pm$0.6 \\ 
$\Teff$ [K] & 10190$\pm$50 \\
$\log{g}$ [cm s$^{-1}$] & 8.00$\pm$0.10 \\
    \hline
     \end{tabular}    
\end{table}       

\begin{figure}
    \includegraphics[width=\columnwidth]{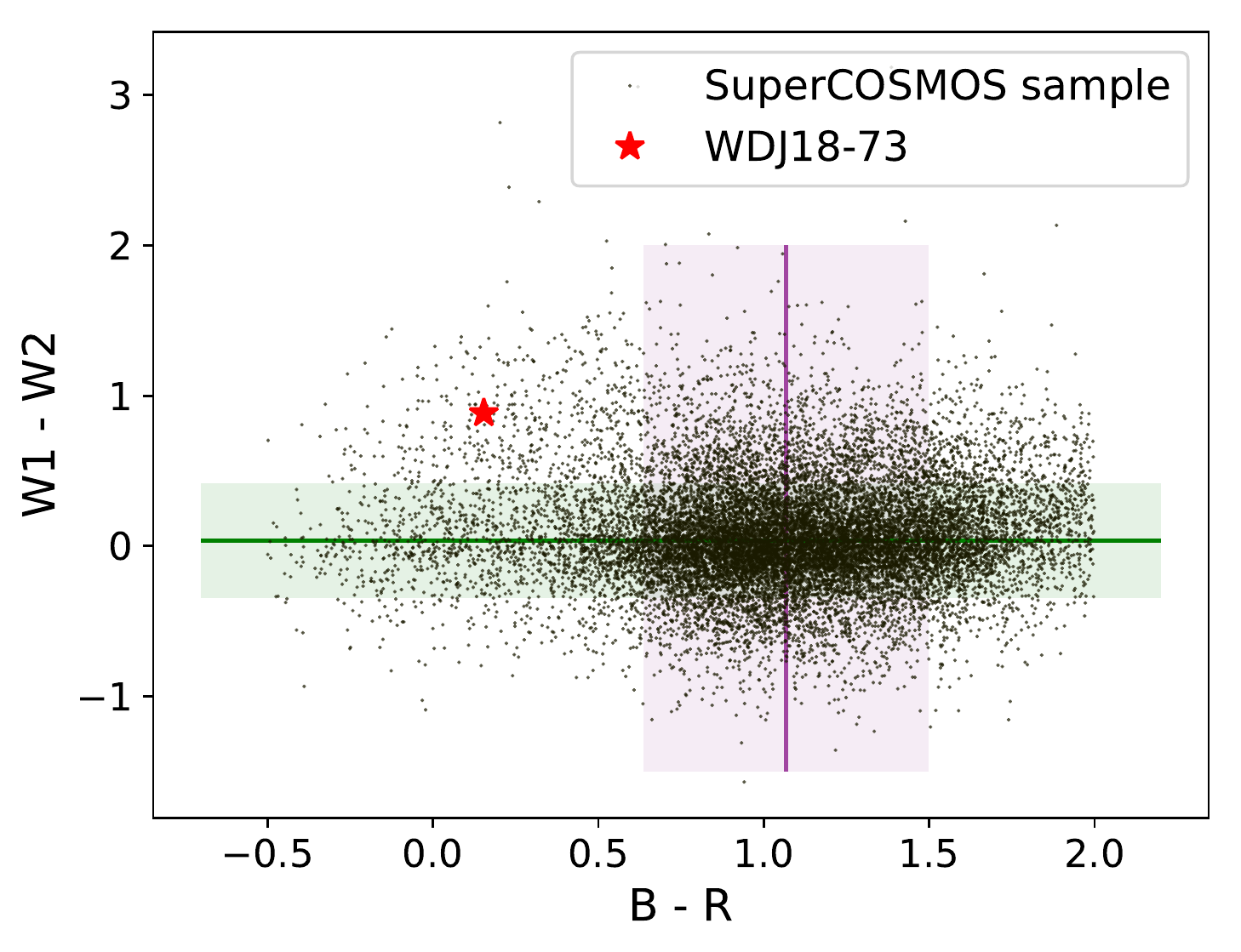}
    \caption{Colour-colour diagram of the sample of 16,928 white dwarf candidates from \citet{DayJones2008} (shown as blue dots) originally selected from SuperCOSMOS that were also present in the \textit{WISE} catalogue. The Y-axis shows the colour in \textit{WISE} bands W1 and W2, and the X-axis the colour in SuperCOSMOS B and R2 bands. The red star indicates the position on this diagram of WDJ1814$-$7354. The purple vertical line shows the position of the distribution's median in the Y-axis, and the 1$\,\sigma$ region is shown as the light purple shaded region. The green horizontal line and region shows the same for the X-axis. }
    \label{fig:colorcolordiscovery}
\end{figure}

WDJ1814$-$7354 was originally identified by one of us (AD) as part of a search for unresolved white dwarf + ultracool dwarf benchmark binaries. The original sample of white dwarf candidates came from a selection of 36,876 objects in the SuperCOSMOS Sky Survey \citep{SuperCOSMOSHambly2001}, following the same selection criteria as in \cite{DayJones2008}. This list was then cross-matched with the All-Sky \textit{WISE} catalogue \citep{Cutri2013} to select targets with infrared excess, that could potentially have an ultracool dwarf companion, obtaining 16,928 objects. WDJ1814$-$7354 is one such outlier in the \textit{WISE} (Wide Field Infrared Explorer, \citeauthor{WISEWright2010} \citeyear{WISEWright2010}) vs SuperCOSMOS colour-colour diagram presented in Fig~\ref{fig:colorcolordiscovery}, lying 2.2$\,\sigma$ away from the median of the distribution of white dwarfs. The \emph{Gaia} Data Release 2 \citep[DR2;][]{Lindegren2018} astrometry, photometry and parameters of WDJ1814$-$7354 are presented in Table \ref{tab:gaiainfo}. The publicly available photometry for WDJ1814$-$7354 is summarized in Table \ref{tab:Photometries}.

\begin{table} 
\begin{minipage}{\columnwidth}

\caption{Apparent magnitudes $m$ of WDJ1814$-$7354 in different bands with central wavelengths $\lambda_c$ in micrometers and the full width at half maximum (FWHM) of the Point Spread Function (PSF) in arcseconds. S-COSMOS is an abbreviation of SuperCOSMOS.}
\label{tab:Photometries}
    \begin{tabular}{lcccc} 
    \hline
    Band (Survey)  & $\lambda_c$ [$\mu$m] & $m$ [mag] &  Epoch & PSF ["] \\ 
    \hline
    B (APASS9) & 0.444 & 16.49$\pm$0.0.11 & 2013.75 & 15 \\ 
    B (S-COSMOS) & 0.467 & 16.345 & 1987.13 & $-$ \\
    g (APASS9) & 0.482 & 16.23$\pm$0.06 & 2013.75 & 15 \\
    G$_{\rm BP}$ (\emph{Gaia}DR2) & 0.513$^1$ & 16.233$\pm$0.005 & 2015.5 & 0.1 \\ 
    V (APASS9) & 0.554 & 16.113$\pm$0.011 & 2013.75 & 15 \\
    R2 (S-COSMOS) & 0.595 & 16.191 & 1987.13 & $-$ \\ 
    r (APASS9) & 0.625 & 16.19$\pm$0.06 & 2013.75 & 15 \\ 
    G (\emph{Gaia}DR2) & 0.640\footnote{Central wavelengths of Gaia passbands are from \citet{Weiler2018}.} & 16.2028$\pm$0.0009 & 2015.5 & 0.1 \\
    G$_{\rm RP}$ (\emph{Gaia}DR2) & 0.778 & 16.104$\pm$0.004 & 2015.5 & 0.1 \\ 
    I (S-COSMOS) & 0.807 & 16.115 & 1987.13 & $-$ \\ 
    J (2MASS) & 1.24 & 15.9$\pm$0.1 & 2001 & 2.5 \\ 
    J (VHS) & 1.252 & 16.078$\pm$0.009 & 2019 & 0.51 \\ 
    H (2MASS) & 1.65 & 15.49$\pm$0.13 & 2013 & 2.5 \\  
    Ks (VHS) & 2.147 & 15.70$\pm$0.03 & 2019 & 0.51  \\ 
    Ks (2MASS)& 2.16 & 15.10$\pm$0.13 & 2013 & 2.5\\
    W1 (\textit{AllWISE}) & 3.35 & 13.94$\pm$0.03 & 2010.59 & 6.1\\ 
    W1 (un\textit{WISE})\footnote{un\textit{WISE} magnitudes were obtained by applying the relations in \citet{Finkbeiner2004} to the fluxes found in the un\textit{WISE} catalogue: 1920$\pm$10 nanomaggies for W1 and 4230$\pm$30 nanomaggies for W2. Systematic uncertainties of 0.02 mag were added \citep[See][]{unwise}} & 3.35 & 14.290$\pm$0.03 & $-$ & 6.1 \\ 
    W2 (\textit{AllWISE}) & 4.60 & 13.07$\pm$0.03 & 2010.59 & 6.4\\
    W2 (un\textit{WISE}) & 4.60 & 13.433$\pm$0.03 & $-$ & 6.4 \\
    W3 (\textit{AllWISE}) & 11.6 & 10.76$\pm$0.09 & 2010.31 & 6.5\\
    W4 (\textit{AllWISE}) & 22.1 & 8.7$\pm$0.4 & 2010.31 & 12 \\
    \hline
    \end{tabular}    
    \end{minipage}
\end{table}

\begin{figure*}
    \includegraphics[scale = 0.6]{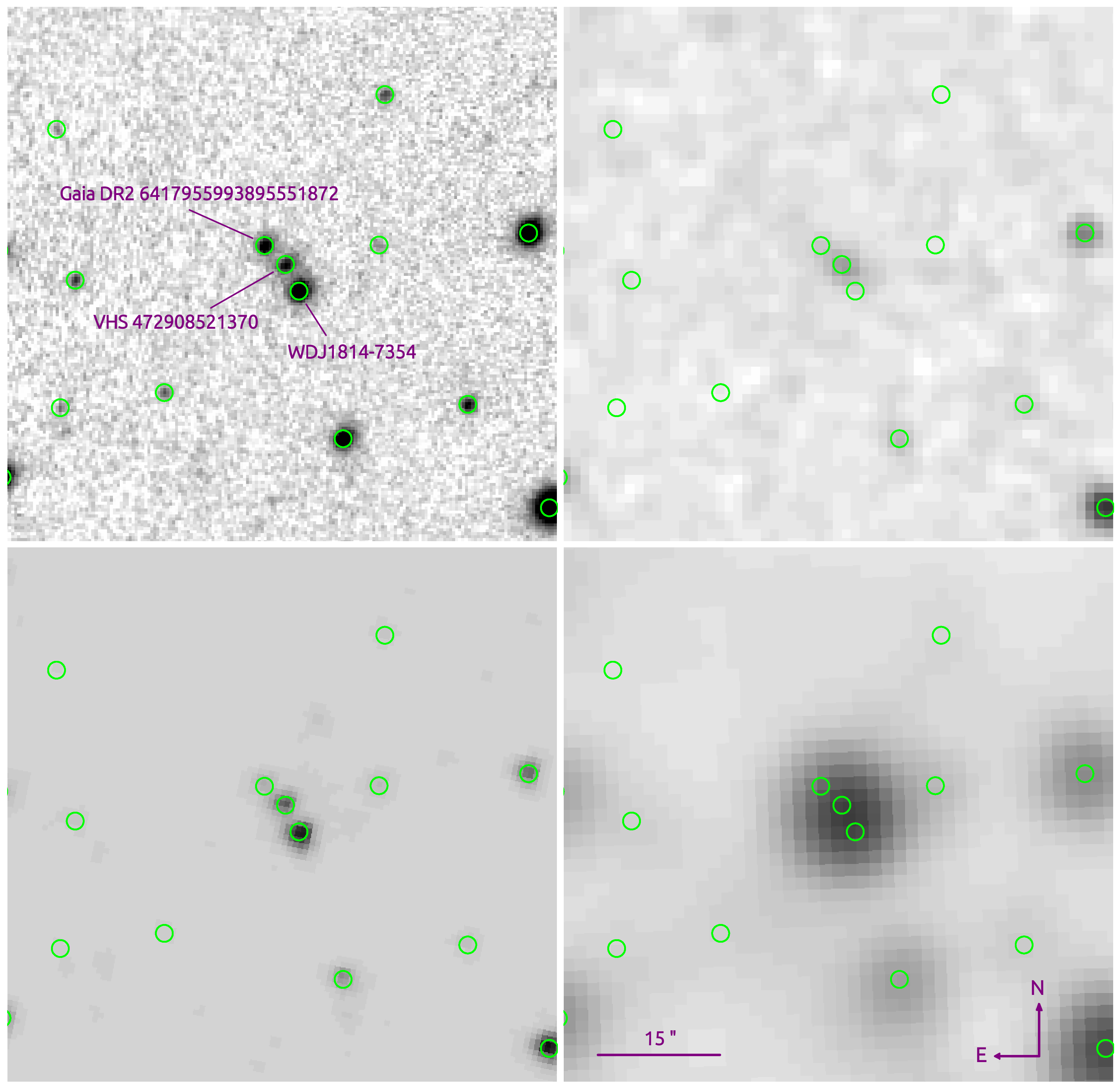}
    \caption{Finding charts in different bands centered on the object WDJ1814$-$7354. From top-left to bottom-right, we display the VHS J-band image, 2MASS J-band image, \textit{Spitzer} IRAC [3.6]-band image and \textit{WISE} W1-band image.  The detected VHS sources are overplotted as green circles, and the orientation and scale are the same in all images. The three sources are labeled in the VHS J image.}
    \label{fig:photometryimages}
\end{figure*}

After comparing images and photometry from the Two Micron All Sky Survey \citep[2MASS;][]{2MASSSkrutskie2006} and VHS (VISTA Hemisphere Survey; \citeauthor{VHS} \citeyear{VHS}) and from un\textit{WISE} and \textit{WISE} (see Fig.~\ref{fig:photometryimages}), we noted that the 2MASS and \textit{WISE} detections are the blend of multiple objects. The \textit{WISE} single source appears as two resolved sources in the un\textit{WISE} catalogue, based on deeper imaging obtained after the coaddition of all 3-5 $\mu$m \textit{WISE} images and with improved modeling of crowded regions \citep{unwise}. Three sources are detected by VHS.

Two objects are identified in \emph{Gaia} DR2, one corresponding to the white dwarf and the other is identified as \textit{Gaia} DR2 6417955993895551872, for which its \textit{Gaia} DR2 information is presented in Table~\ref{tab:gaiainfobackground}. VHS and un\textit{WISE} photometry for this source and the third source, identified as VHS 472908521370, is presented in Table \ref{tab:Photometries3}. VHS 472908521370 has no counterpart in \textit{Gaia} DR2. The angular separation from WDJ1814$-$7354 is $\sim$3.2 arcseconds for VHS472908521370 and $\sim$6.3 arcseconds for \textit{Gaia} DR2 6417955993895551872. These additional objects are likely contributing to the 2MASS and \textit{WISE} IR excess, but their contribution to optical photometry is negligible: \textit{Gaia} DR2 6417955993895551872 is 3.5 mag fainter in the G-band with respect to WD1814$-$7354. VHS 472908521370 is even fainter, given its non-detection in \textit{Gaia} DR2, which combined with its VHS colours indicate a very red object. 

\begin{table} 
\centering
\caption{\textit{Gaia} DR2 astrometry and photometry of the object \textit{Gaia} DR2 6417955993895551872, along with its distance (d) in parsecs from \citet{BailerJones2018}. Barycentric right ascension ($\alpha$) and declination ($\delta$) are in the International Celestial Reference System and at \textit{Gaia} DR2 reference epoch 2015.5. Proper motion in right ascension ($\mu_{\alpha}\cos{\delta}$) and declination ($\mu_{\delta}$) and parallax ($\varpi$) are also at \emph{Gaia} epoch, 2015.5. Apparent magnitudes are in the three \emph{Gaia} DR2 passbands. }
\label{tab:gaiainfobackground}
    \begin{tabular}{lc} 
    \hline
    Gaia Source ID  & \textit{Gaia} DR2 6417955993895551872 \\
    \hline
$\alpha$ [deg] & 273.57335015736 \\  
$\delta$ [deg] &  $-$73.91602610506 \\
$\mu_{\alpha}\cos{\delta}$ [mas yr$^-1$] & $-$7.4$\pm$0.1 \\ 
$\mu_{\delta}$ [mas yr$^-1$] & 4$\pm$1\\ 
$\varpi$ [mas] & $-$1.8$\pm$0.8\\
G [mag] &  19.877$\pm$0.008 \\
G$_{\rm BP}$ [mag] &  20.62$\pm$0.08 \\
G$_{\rm RP}$ [mag] &  18.67$\pm$0.03 \\
G$_{\rm BP}$ $-$ G$_{\rm RP}$ [mag] &  1.95$\pm$0.11 \\
d [pc] & 3200$^{+5000}_{-1800}$ \\
    \hline
     \end{tabular}    
\end{table}       

\textit{Gaia} DR2 6417955993895551872 is likely a distant background object, as its proper motion is very small and not compatible with that of WDJ1814$-$7354, and its distance is not well constrained. This object has also photometric information in VHS catalog but not in the un\textit{WISE} catalog.

\begin{table} 
\caption{VHS and un\textit{WISE} apparent magnitudes and colour of the two other sources identified around the position of WDJ1814$-$7354.}
\label{tab:Photometries3}
    \begin{tabular}{lccc} 
    \hline
    Band (survey) & VHS 472908521370 & \textit{Gaia} DR2 \\
    & & 6417955993895551872  \\
    \hline
    J (VHS) [mag] & 17.78$\pm$0.04 & 17.263$\pm$0.025  \\
    Ks (VHS) [mag] &  16.18$\pm$0.04 & 16.52$\pm$0.06 \\
    J $-$ Ks [mag] & 1.6$\pm$0.08 & 0.74$\pm$0.08  \\
    W1 (un\textit{WISE}) [mag] & 15.144$\pm$0.011 & $-$  \\
    W2 (un\textit{WISE}) [mag] & 14.643$\pm$0.024 & $-$  \\
    W1 $-$ W2 [mag] & 0.50$\pm$0.03 & $-$  \\
    \hline
    \end{tabular}    
\end{table}      

\section{Follow-up observations}
\label{sec:observations}

\subsection{X-SHOOTER spectroscopy}
\label{sec:X-Shooter}

\subsubsection{Observations in 2011}

WDJ1814$-$7354 was observed on 2011 September 18 with the multi-wavelength X-SHOOTER spectrograph \citep{Vernet2011}, mounted at the Cassegrain focus of the Very Large Telescope (VLT) UT2 in Cerro Paranal (Chile), as part of the ESO programme ID 087.C$-$0639B. The observing mode was SLITSPEC 100k/1pt/hg, longslit spectroscopy with auto-nodding along the slit. The three X-SHOOTER arms covering the ultraviolet (UVB, 3000$-$5595 \AA), visible (VIS, 5595$-$10240 \AA) and near-infrared (NIR, 10240$-$24800 \AA) parts of the spectrum were used to take four subsequent exposures of 500 seconds each (590 in the case of the NIR arm), giving 2000 seconds of total exposure time for the UVB and VIS arms and 2360 seconds for the NIR arm. We used narrow slits of 1.0" width for the UVB arm and 0.9" for the VIS and NIR arms, giving spectral resolutions $R=\lambda/\Delta\lambda$ of 5400 (UVB), 8900 (VIS) and 5600 (NIR). The signal-to-noise ratio (SNR) achieved was around 40 for the VIS and 20 for the UVB part of the spectrum, enough for precise model fitting. The NIR part did not contain useful information for white dwarf modelling or abundance analysis. The weather conditions of the observing night were good, with seeing around 0.5$-$0.6", and the airmass was 1.5$-$1.6 during the observations of this target.

\subsubsection{Observations in 2019}

WDJ1814$-$7354 was observed again in 2019 with X-SHOOTER as part of the ESO programme ID 0103.C-0431(B), with PI S. Xu. Stare mode was used, and the exposure times are as follows: 2 exposures of 1700 seconds with the UVB arm (total of 3400 seconds) and 2 exposures of 1729 seconds with the VIS arm (3458 seconds total). The NIR part was not needed, so just 1 exposure of 100 seconds was taken. The signal-to-noise ratio (SNR) achieved was around 35 for the VIS and 20 for the UVB part of the spectrum. The weather conditions of the observing night were excellent, with seeing around 0.4$-$0.5", and the airmass was around 1.6$-$1.7 during the observations.

To reduce both spectra we used the ESO X-SHOOTER reduction pipeline \citep{Xshooterpipeline} and the software {\sc{esoreflex}}\footnote{Available at \url{https://www.eso.org/sci/software/esoreflex/}}. 

\subsection{\textit{Spitzer} observations}
\label{sec:spitzer}

WDJ1814$-$7354 was observed with \textit{Spitzer} on 2019 July 10 as part of the DDT program ID 14220, with PI Dr. Siyi Xu (see \citeauthor{XuLai2020} \citeyear{XuLai2020}), with the instrument and mode IRAC/Map PC, in both bands IRAC [3.6] and [4.5], centered at wavelengths 3.6 and 4.5 microns, respectively. The observing strategy was 30 seconds frame time with 11 medium size dithers for each wavelength. 

The \textit{Spitzer} [3.6] band image of WDJ1814$-$7354 is shown in Fig.~\ref{fig:photometryimages} (bottom-left panel). Three sources are clearly distinguished in the Spitzer images. The relative fluxes of two of these sources (at the positions of WDJ1814$-$7354 and VHS 472908521370) were extracted after performing aperture photometry on successfully PSF-fitted sources positions using the stack of images, following Recipe 7 of the IRAC Instrument Handbook\footnote{Available at  \urlx{https://irsa.ipac.caltech.edu/data/SPITZER/docs/irac/iracinstrumenthandbook/IRAC_Instrument_Handbook.pdf}} adapted to the command-line {\sc{mopex}} \citep{mopex}. Passive deblending of the sources was automatically performed with {\sc{mopex}} routines, to take contamination from nearby sources into account. The uncertainty in Spitzer fluxes was the sum between the uncertainty obtained after the MOPEX extraction and the median values of the rms variatins due to the relative repeatability of IRAC reported by \citet{Reach2005b} of 1.7 per cent for channel 1 and 2.2 per cent in channel 2.

The \textit{Spitzer} fluxes were later corrected following section 7.9 of the IRAC Instrument Handbook. The first correction was applied by dividing the PRF fluxes by the correction factors in table C.1 of the IRAC Instrument Handbook, whose values are 1.021 for IRAC1 and 1.012 for IRAC2. The second correction mentioned was not applied, as this correction applies to blue sources, being null for red sources like our target.

Apparent magnitudes were computed using Vega zero flux values determined with the Python \textit{Pyphot} package, that agree with the zero fluxes listed in the IRAC Instrument Handbook (280.9$\pm$4.1 Jy for IRAC [3.6] and 179.7$\pm$2.6 Jy for IRAC45). The results are presented in Table~\ref{Spitzerphotometry}. 

\begin{table} 
\centering
\caption{\textit{Spitzer} extracted photometry for WDJ1814$-$7354 and VHS 472908521370. $f_{36}$ and $f_{45}$ are the fluxes in \textit{Spitzer} IRAC [3.6] and IRAC [4.5] bands in $\mu$Jy units obtained using MOPEX and $m_{36}$ and $m_{45}$ are the apparent magnitudes computed from these fluxes.  }
\label{Spitzerphotometry}
    \begin{tabular}{lcc}   
    \hline
        & WDJ1814$-$7354 & VHS 472908521370 \\ 
    \hline
$f_{36}$ [$\mu$Jy] & 591$\pm$7 & 297$\pm$5 \\  
$f_{45}$ [$\mu$Jy] & 643$\pm$9 & 353.2$\pm$6 \\
$m_{36}$ [mag] & 14.185$\pm$0.013 & 14.933$\pm$0.017 \\
$m_{45}$ [mag] & 13.622$\pm$0.015 & 14.272$\pm$0.017 \\
    \hline
     \end{tabular}    
\end{table}

\subsection{FIRE spectroscopy}

We explored the possibility of VHS 472908521370 being a cool companion of WDJ1814$-$7354. If we place it at the same distance as WDJ1814$-$7354, its absolute magnitudes in J and Ks VHS bands and in \textit{Spitzer} bands would be consistent with a brown dwarf of spectral type L5.5-L6, according to the relations from \citeauthor{DupuyLiu2012} (2012; Table 14). To discard or confirm this possibility, we obtained a spectrum of this object with the Folded-port InfraRed Echellette (FIRE, \citeauthor{FIRE} \citeyear{FIRE}) at the Magellan Baade telescope, Las Campanas observatory, on the night of June 02 2019. We used the low-resolution prism mode, with a 0.6" wide slit, to obtain 8 exposures with a total integration time of 1732 sec (28,8 min). We used the typical ABBA nodding pattern. 

We used the {\sc firehose} pipeline \citep{firehose} to reduce the spectra which include: flat fielding, wavelength calibration, A-B pair subtraction to remove first-order sky emission. We then trace the spectra in each positive and negative feature and remove the residual background emission using the flux that fell on the slit. We then combined the 8 individual spectra using a robust weighted mean. For telluric correction and flux calibration we used the A0V star  HD 167061  and we reduced following an identical procedure  than for the object and to construct and apply the telluric correction we used the IDL based task {\sc xtellcor} \citep{Vacca2003}.

\begin{figure}
    \includegraphics[width=\columnwidth]{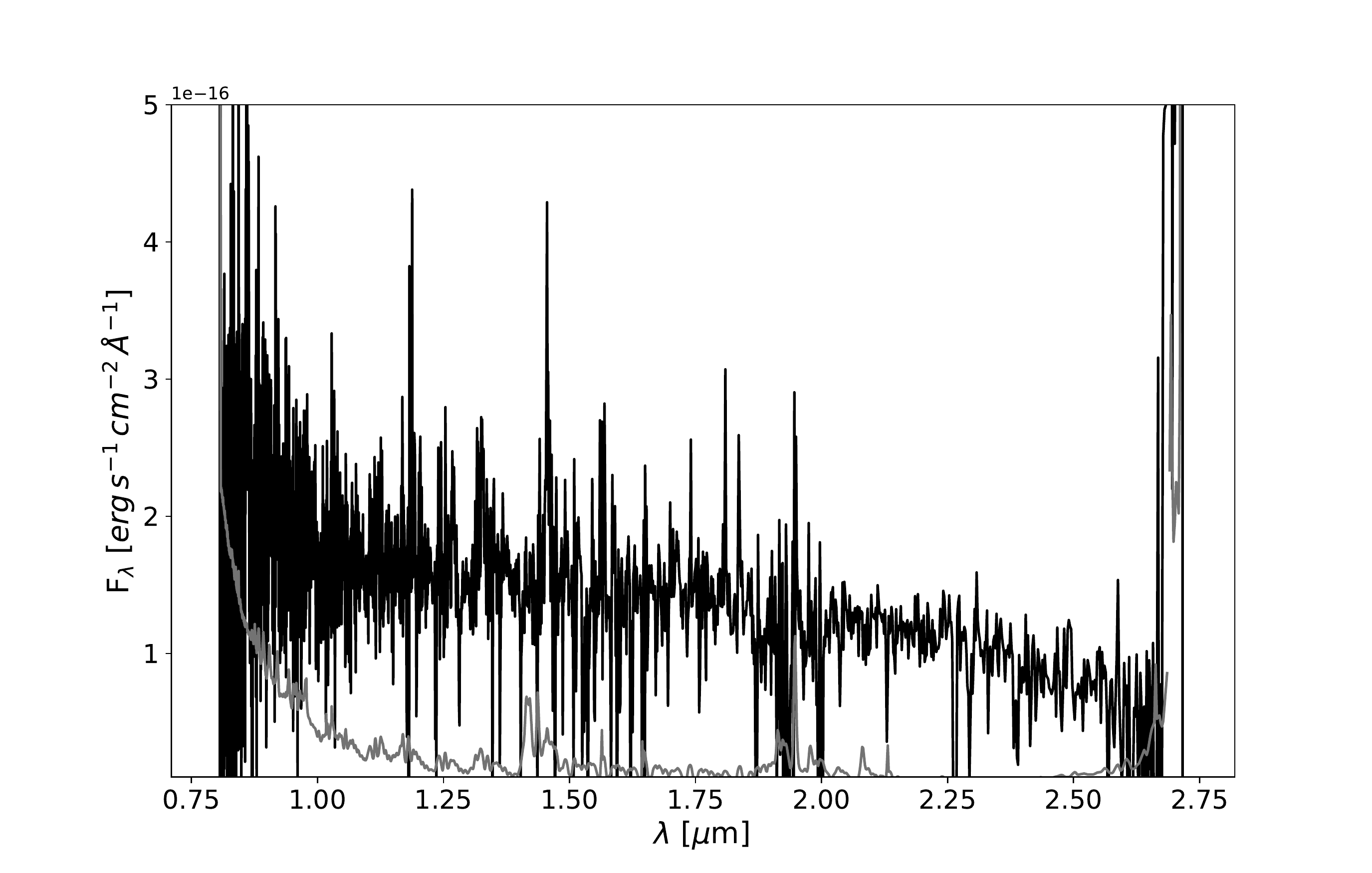}
    \caption{Spectrum of VHS 472908521370 obtained on the night of June 02 2019 with the FIRE instrument. In black it is shown the combination of the individual spectra using a robust weighted mean, and in grey the formal error in flux.}
    \label{fig:firespectrum}
\end{figure}

The FIRE spectrum is shown in figure \ref{fig:firespectrum}. Although the signal-to-noise ratio is $\sim$1.9, it does not show any evident characteristic features of substellar objects, such as strong water bands, iron hydrates (FeH), NaI, CO or methane absorption. We therefore conclude that VHS 472908521370 is likely another background object. It is brighter in the infrared and closer to the white dwarf than \textit{Gaia} DR2 6417955993895551872, and it is likely contributing to 2MASS and \textit{WISE} photometry.

\section{Analysis}
\label{sec:analysis}

\subsection{White dwarf properties}
\label{sec:spectroscopicfit}

\begin{figure}
    \includegraphics[width=\columnwidth]{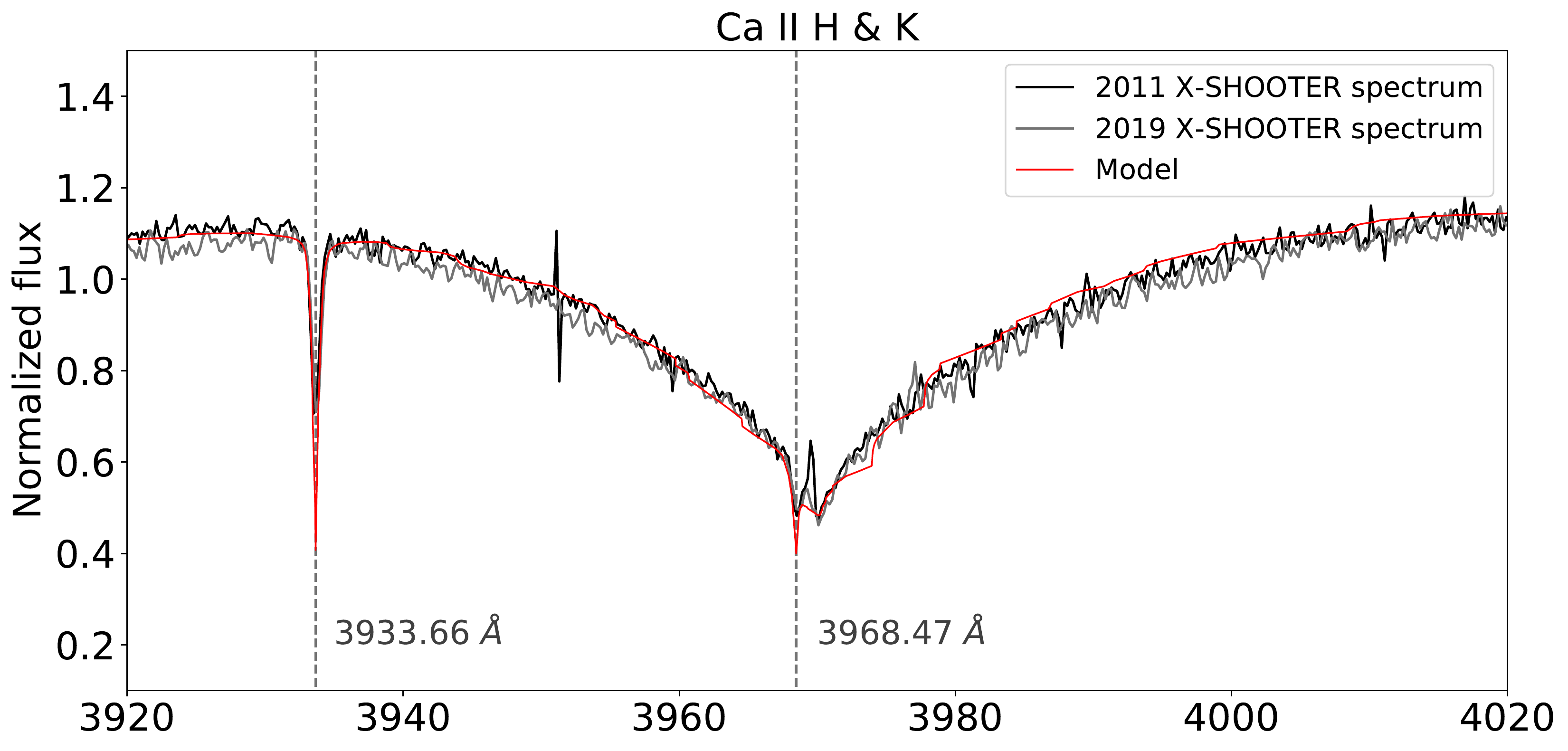}
    \includegraphics[width=\columnwidth]{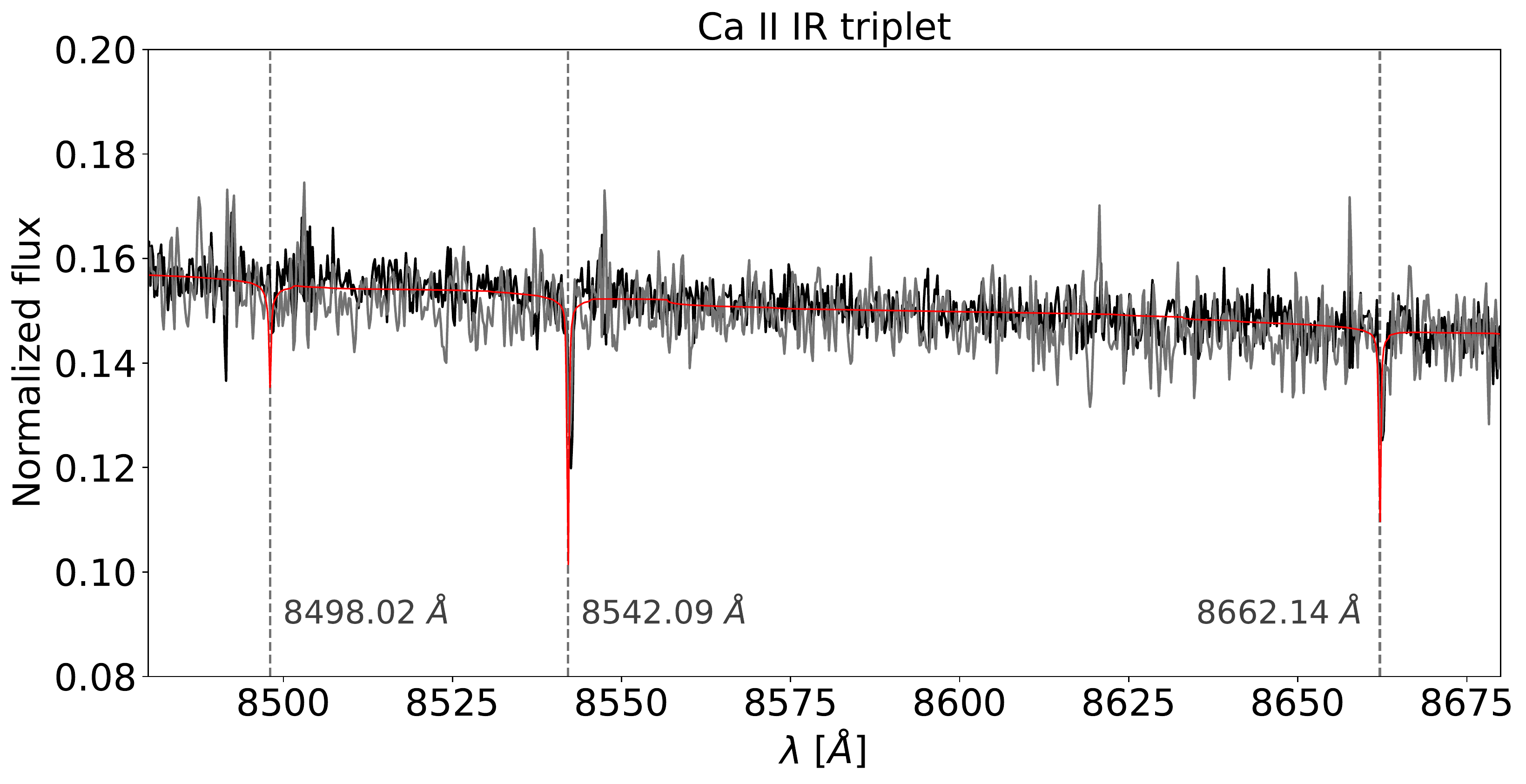}
    \caption{Calcium absorption lines found in WDJ1814$-$7354 X-SHOOTER 2011 (black) and 2019 (grey) spectra, marked with grey vertical dashed lines. The white dwarf model is superimposed in red. Line wavelength values from NIST Atomic Spectra Database \citep{NIST}.
    }
    \label{fig:metal_lines}
\end{figure}

\begin{figure}
	\centering
	\includegraphics[width=\columnwidth]{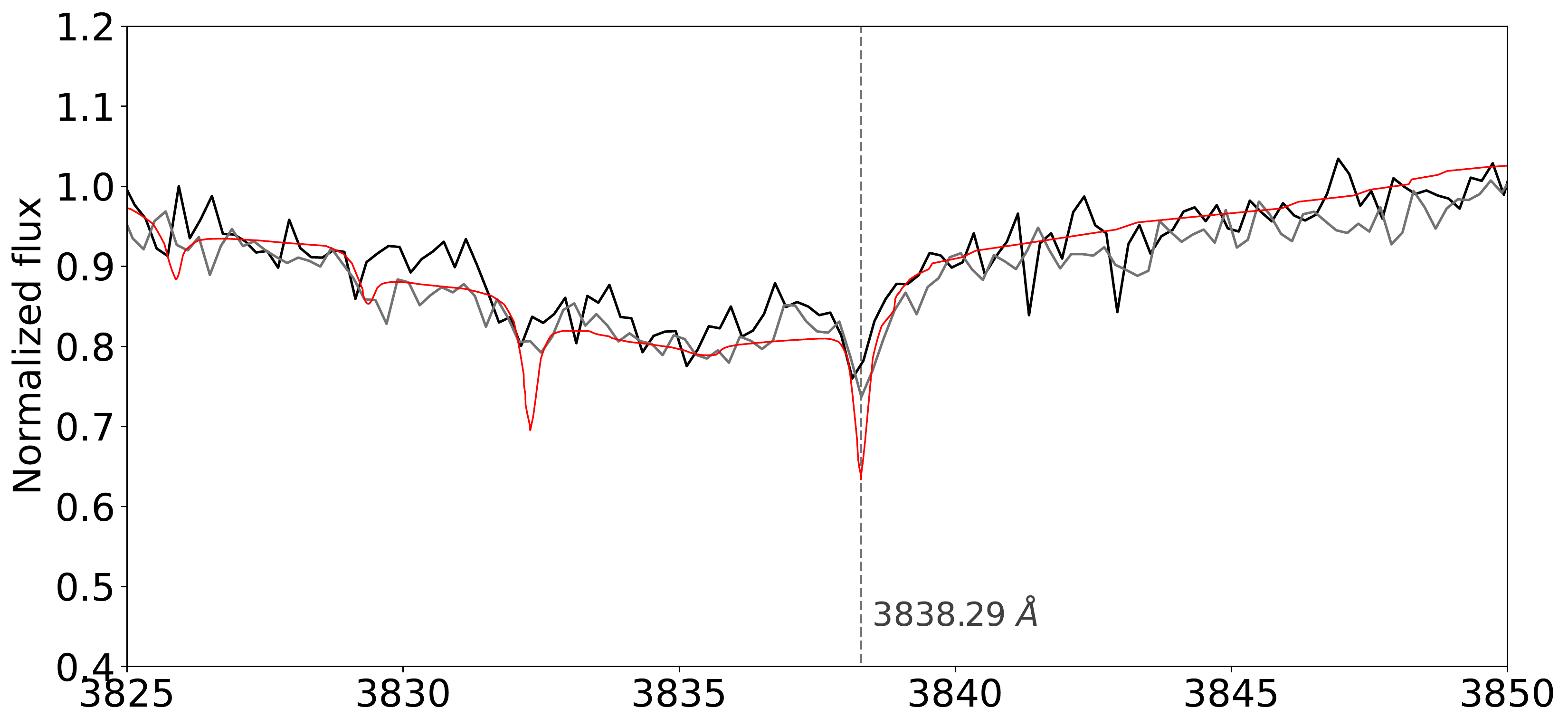} 
	\includegraphics[width=\columnwidth]{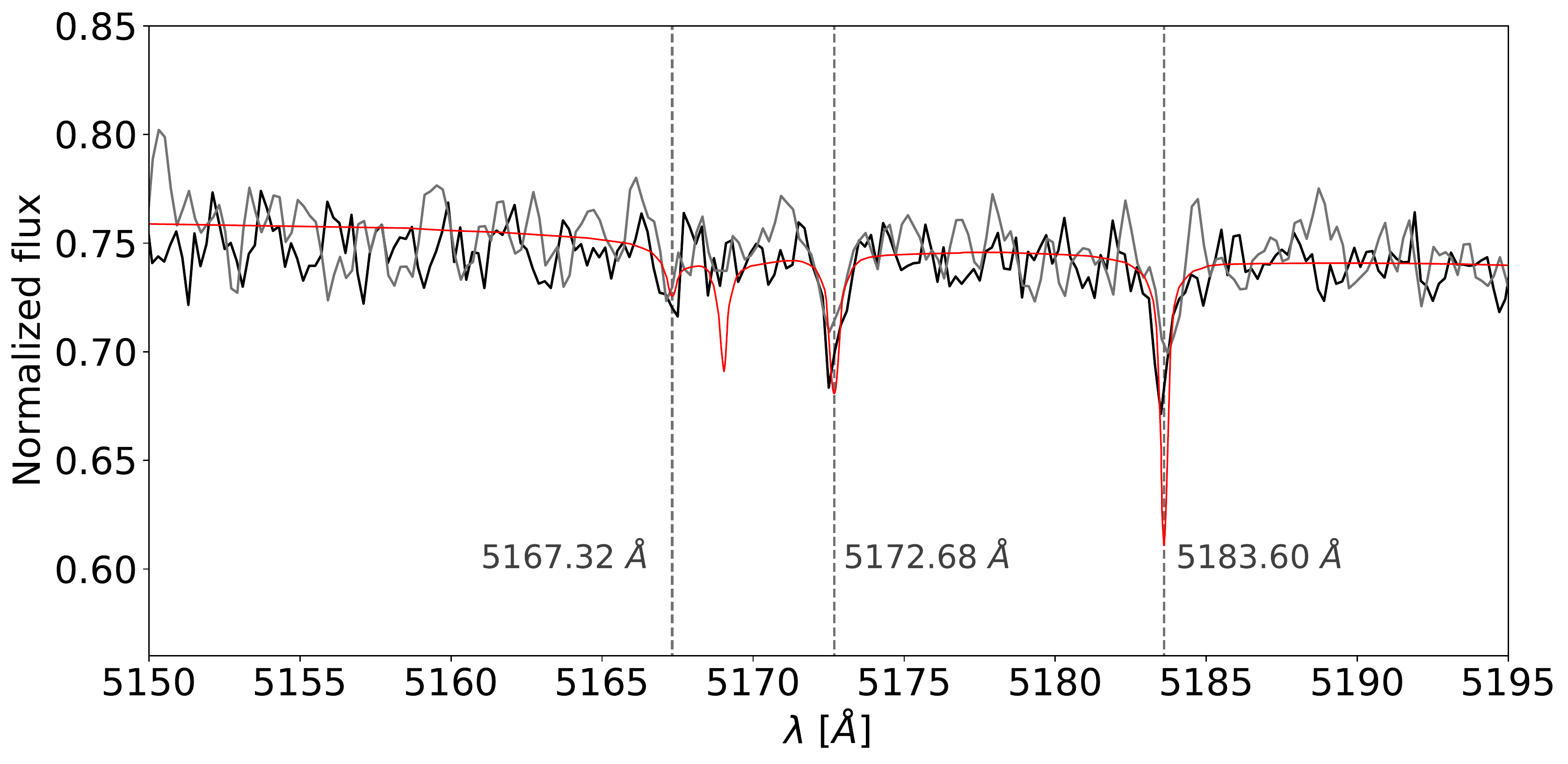}
	\caption{Mg I absorption lines found in WDJ1814$-$7354 X-SHOOTER 2011 (black) and 2011 (grey) spectra, marked with grey vertical dashed lines. The white dwarf model is superimposed in red. Line wavelength values from NIST Atomic Spectra Database \citep{NIST}.
	}
	\label{fig:metal_lines2}
\end{figure}

The spectrum of WDJ1814$-$7354 is that of a typical DAZ white dwarf, characterised by broad Balmer lines and absorption from Mg\,{\sc i--ii} and Ca\,{\sc i--ii} species and weaker absorption lines from Fe\,{\sc i}. Examples of Mg and Ca lines in both 2011 and 2019 spectra are shown in Figs \ref{fig:metal_lines}--\ref{fig:metal_lines2}. Fe lines are not shown in the figures due to their faintness, but with a large number of them and consistent results we are confident of the Fe detection. 

We analysed the combined UVB and VIS 2011 spectra via comparison with a grid of synthetic spectra modelled with the stellar atmosphere code outlined by \citet{Koester2010, Koester2020}. The NIR spectrum was not used due to its low SNR. The 2019 spectrum was used to investigate the potential variability of metal absorption lines within the 1-sigma errors (see Section \ref{sec:abundances}).

We started by fitting the full Balmer series, incrementally including the identified pollutants after finding the initial best fit and subsequently adjusting the result. The best-fit is obtained for $T_{\rm eff} = 10,140$\,K and $\log{g} = 8.10$. Statistical uncertainties are negligible due to the high SNR of the optical spectrum. 
Subsequently, we performed a photometric fit of the available APASS and {\em Gaia} DR2 photometry (Table\,\ref{tab:Photometries}), in which we used the {\em Gaia} DR2 parallax and the white dwarf mass-radius relation \citep{Fontaine2001} as additional priors. We have not used VHS photometry to be sure that the IR excess has no effect in the fit. The photometric fit delivers a 0.1\,dex-lower surface gravity. Hence, we iterated between spectroscopic ($\log{g}$ fixed) and photometric ($\Teff$ fixed) fits, until both methods converged to a final solution. The difference between the initial spectroscopic fit and the final result is adopted as the error estimate, resulting in $T_{\rm eff} = 10,090 \pm 50$\, K and $\log{g} = 8.00 \pm 0.13$. The interstellar reddening obtained as a result of our fitting procedure is negligible, and of the order of 0.01\,mag.

We measured the radial velocities of the strongest metal lines via fitting gaussian models and estimate the uncertainty as the scatter between the measurements of the different lines. For the 2011 spectrum we obtained an average radial velocity of $\overline{\rm RV}_{2011}$ = 44$\pm$5 km s$^{-1}$ and for the 2019 spectrum, $\overline{\rm RV}_{2019}$ = 47$\pm$8 km s$^{-1}$. The individual radial velocities for the different lines are presented in table \ref{RVs}.

The atmospheric parameters of WDJ1814$-$7354 correspond to a white dwarf mass of $0.59 ^{+0.08}_{-0.06}$\,M$_{\sun}$ and a cooling age of $\tau_{\rm cool} = 0.58 ^{+0.14}_{-0.07}$\,Gyr. Using the \citet{Cummings2018} initial-to-final-mass relation (IFMR), we infer a progenitor mass of $1.3^{+0.5}_{-0.4}$\,M$_{\sun}$. To estimate the progenitor age we used the evolutionary tracks of \citet{Choi2016}. For a progenitor mass of 1.3 M$_{\sun}$ (late F-type star), the progenitor age is $\tau_{\rm MS} = 5.2$\,Gyr, and the total age of the white dwarf would be $\tau_{\rm MS}$ + $\tau_{\rm cool}$ = 5.8 Gyr.
The large systematic uncertainty implied by the empirical IFMR translates into a large uncertainty on the progenitor age. For the upper limit of the progenitor mass of 1.8 M$_{\sun}$ (a late A-type star), the progenitor age would be $\tau_{\rm MS} = 3.4$\,Gyr, so the lower limit on the total age of the white dwarf is 4 Gyr. The upper limit on the total age is unconstrained, because the lower limit of the progenitor mass is 0.9 M$_{\sun}$ (an early K dwarf star), whose lifetime could be comparable or longer than the age of the Universe.

\begin{table} 
\centering
\caption{Radial velocities computed in both 2011 and 2019 X-SHOOTER spectra. The average values are $\overline{\rm RV}_{2011}$ = 46$\pm$8 km s$^{-1}$ and $\overline{\rm RV}_{2019}$ = 45$\pm$10 km s$^{-1}$. }
\label{RVs}
    \begin{tabular}{lcc}   
    \hline
        Line (Wavelength) [\AA] & RV$_{2011}$ [km s$^{-1}$]  & RV$_{2019}$ [km s$^{-1}$] \\ 
        
    \hline
        Ca {\sc ii} (3933.66) & 46 & 54 \\ 
        Ca {\sc ii} (3968.47) & 58 & 54 \\ 
        Ca {\sc ii} (8662.14) & 47 & 31 \\ 
        Ca {\sc i} (4226.73) & 39 & 48 \\ 
        Mg {\sc i} (3838.29) & 41 & 52 \\ 
        Mg {\sc i} (5167.32) & 42 & 41 \\ 
        Mg {\sc i} (5172.68) & 46 & 40 \\  
        Mg {\sc i} (5183.69) & 43 & 56 \\ 

    \hline
     \end{tabular}    
\end{table}       

\subsection{Composition of the accreted material and diffusion timescales}
\label{sec:abundances}

The abundances of the three detected elements (Mg, Ca, and Fe) are given in Table\,\ref{tab:abundances}. The listed  abundances and uncertainties are the averages and the standard deviations inferred from the modelling of the strongest lines for each detected element: lines at 3933.66, 3968.47, 8498.02, 8542.09, 8662.14 (Ca {\sc ii}) and 4226.73 \AA\, (Ca {\sc i}) for Calcium; lines at 4481.13 (Mg {\sc ii}), 3838.29, 5167.32, 5172.68 and 5183.69 \AA\, (Mg {\sc i}) for Magnesium; and lines at 3609, 3619, 3632, 3720, 3738, 3750, 3759, 3816, 3821 and 3826 \AA\, for iron (Fe {\sc i}). In addition, we derive upper limits for Na, Al, Si, P, S, Ti, and Ni. 

Using the accretion-diffusion models of \citet{Koester2009}, we estimate a mass-fraction contained within the convection zone of $\log{M_{\rm cvz}/M_{\rm wd}} = -11.1$. The diffusion time-scales of the three detected pollutants are of the order of 10--20\,yr. Given these relatively short diffusion timescales, we tested whether any measurable changes occurred during the $\approx 10$-yr baseline between our first observation and the more recent 2019 spectrum. For the three strongest lines of CaII and MgI there is an indication that the equivalent widths of 2019 spectrum are smaller by 3$-$6 per cent than in 2011. The lines have the same widths, but are slightly less deep, which might be caused by a small difference in resolution of the two spectra. This difference is marginally compatible with the mutual errors. For the smaller lines, any differences are within the errors. If the differences are real, they would correspond to differences in the abundances by about 0.02-0.04 dex, within the errors of our determinations. We interpret this result as an indication of WDJ1814$-$7354 currently accreting from its debris disc. 

In Table\,\ref{tab:abundances}, we list the diffusion timescales,  defined as the time necessary for the flux to be reduced by 1/e, corresponding to the model atmosphere of  WDJ1814$-$7354 and the accretion fluxes we measure for each detected elements as well as those with upper limits. The accretion flux is defined as $\dot{\rm M} = M_{\rm wd} \times 10^{q} \times A \times 10^{[\rm Z/H]} / \tau_d$, where $M_{\rm wd}$ is the mass of the white dwarf, $A$ is the atomic weight  and $ q = \log_{10} (M_{\rm CVZ} / M_{\rm wd})$, being $M_{\rm CVZ}$ the mass of the convection zone. The systematic errors from the parameters on the accretion fluxes largely cancel when we use abundance ratios. 

\begin{table}
\centering
\caption{Metal abundances measured in WDJ1814$-$7354 X-SHOOTER spectrum from 2011. [Z/H] $ = \log n(\rm Z)/n(\rm H)$ is the abundance of the element Z, $\tau_d$ is the diffusion timescale and $\dot{\rm M}$ is the accretion flux. The total accretion flux includes only the observed elements, not the upper limits. }
\label{tab:abundances}
    \begin{tabular}{lccc}  
    \hline
    Element & [Z/H]  & $\tau_d$ [yrs] & log|$\dot{\rm M}$| [g s$^{-1}$] \\ 
    \hline
    Na &  <$-$7.40 &  22.42 &  <7.124 \\ 
    Mg &  $-$6.14$\pm$0.08  &  21.66 &  8.42$\pm$0.08 \\ 
    Al &  <$-$7.30 &   19.78 &  <7.35 \\ 
    Si &  <$-$6.00  & 19.32 &  <8.67 \\
    P &   <$-$4.50 & 17.701 &  <10.26 \\
    S &   <$-$4.50  & 17.33 &  <10.28 \\
    Ca &  $-$7.22$\pm$0.15 & 14.415 & 7.74$\pm$0.15 \\
    Ti &  <$-$8.00 &  12.21 &  <7.12 \\
    Fe &  $-$6.06$\pm$0.19  & 10.814 & 9.17$\pm$0.19 \\
    Ni & <$-$6.30  &  10.44 &  <8.96 \\
    Total & $-$ & $-$ & 9.25$\pm$0.17 \\
    \hline
        \end{tabular}    
\end{table}

\subsection{Disc model fit}
\label{sec:discfit}

\begin{figure*}
    \includegraphics[scale=0.6]{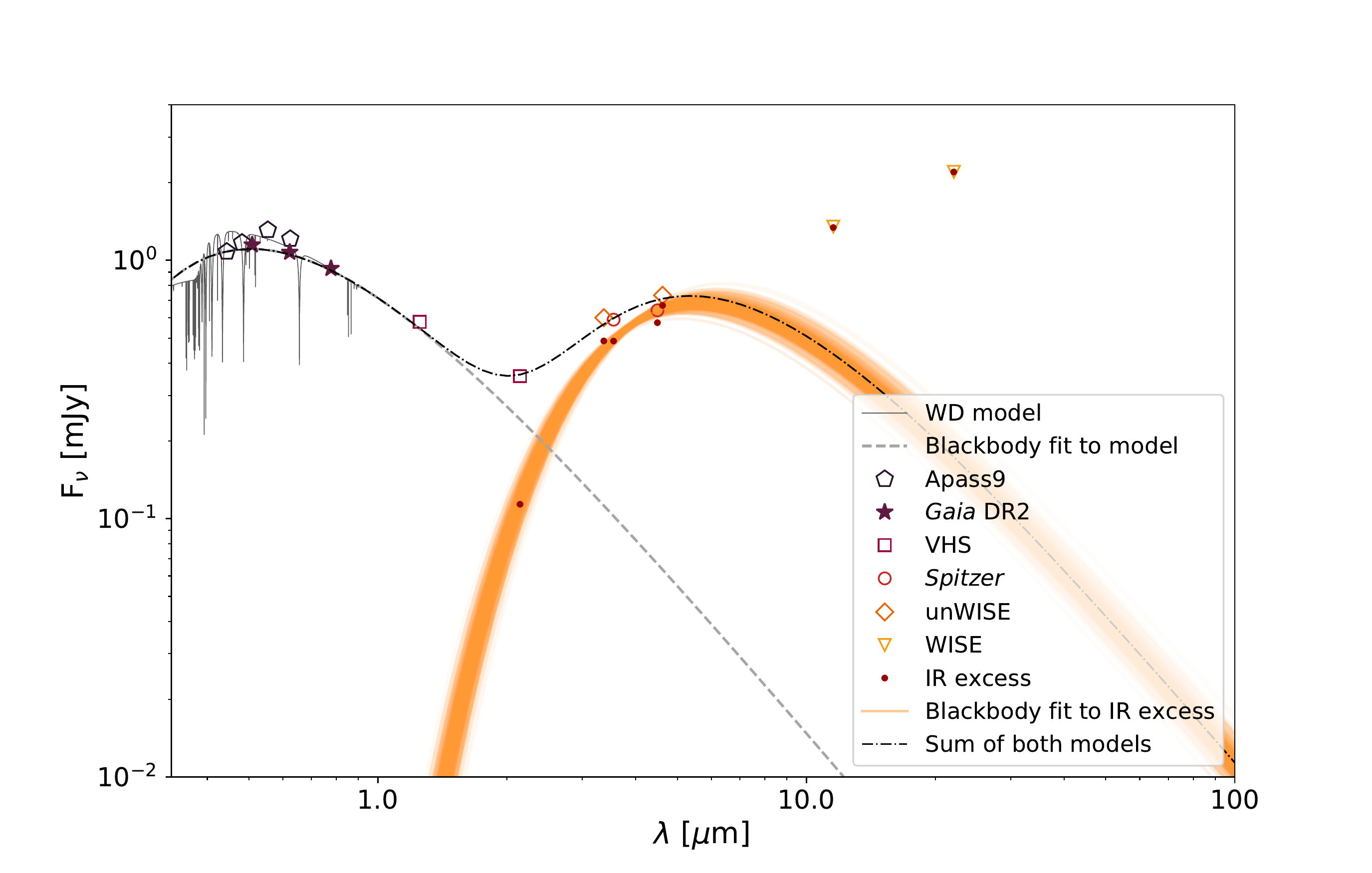}
    \caption{WDJ1814$-$7354 spectral energy distribution and IR excess along with a random set of 500 blackbody fits to the IR excess and the sum of the disc model (blackbody fit to the median of the parameter distribution) and the white dwarf backbody model (dashed-dotted line) expressed in flux density units (mJy). The \textit{WISE} fluxes are upper limits, as WDJ1814$-$7354 photometry is blended with the background sources in \textit{WISE} bands. 
    }
    \label{fig:disc_bbfit}
\end{figure*}

The infrared excess of WDJ1814$-$7354 detected by 2MASS and \textit{WISE} persists in the higher-resolution photometry of VHS, un\textit{WISE} and \textit{Spitzer} (see Fig. \ref{fig:disc_bbfit}).

We have estimated the IR excess as the difference between a blackbody fit to the white dwarf model scaled to \textit{Gaia} band G$_{\mathrm{RP}}$ photometry and the resolved photometry (VHS, un\textit{WISE} and \textit{Spitzer}). We have modelled this IR excess with a simple blackbody function:

\begin{equation}
    B_{\lambda} = \alpha \frac{2 \pi c^2}{\lambda^5}\frac{1}{e^{\frac{hc}{\rm K_B T_{bb}\lambda}} - 1} 
	\label{eq:blackbody}
\end{equation}

where $h$ is the Plank constant, $c$ the speed of light, K$\rm_B$ the Boltzmann constant, T$\_{bb}$ the temperature of the blackbody and $\alpha$ is the scaling factor.

We have used a blackbody model instead of a Flat Ring Model because we lack reliable photometry for wavelengths longer than 4.5 $\mu$m, necessary to properly constrain the inner and outer temperatures of the disc. We note the limitations of using a single temperature blackbody to model the dust and that more detailed modelling is required to fully constrain the properties of the disc.

For computing the parameters of the disc blackbody model fit ( temperature T$_{\rm bb}$ of the disc and scaling factor $\alpha$) and their uncertainties, we have produced a set of 10,000 simulated random data points drawn from a Gaussian distribution, with mean $\mu$ equal to each IR excess value (excluding the blended 2MASS and \textit{WISE} photometry) and $\sigma$ equal to its corresponding error, obtained after a linear propagation of the errors from the photometry and the previous blackbody fit to the white dwarf model.

\begin{figure}
    \includegraphics[width=\columnwidth]{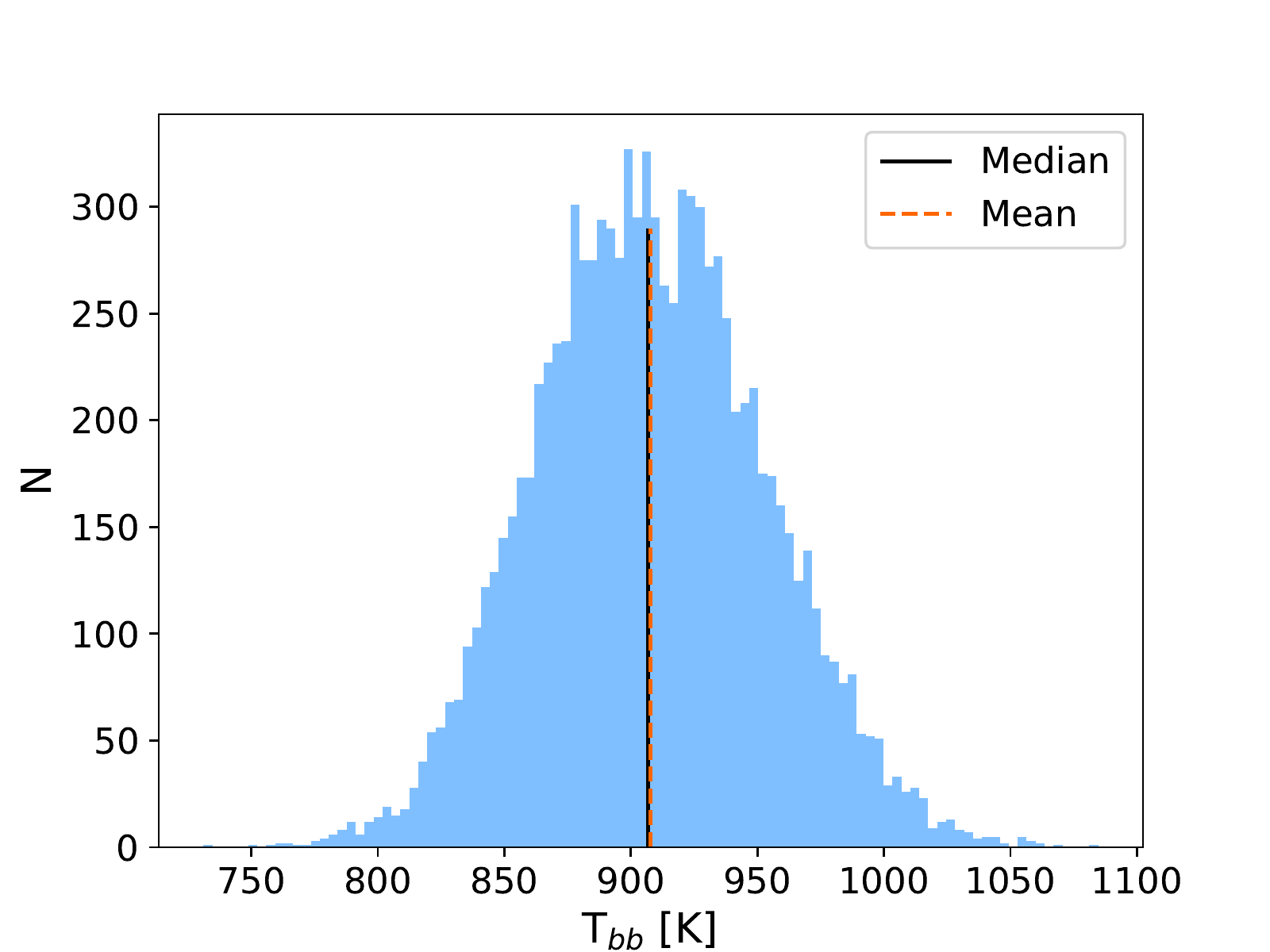}
    \includegraphics[width=\columnwidth]{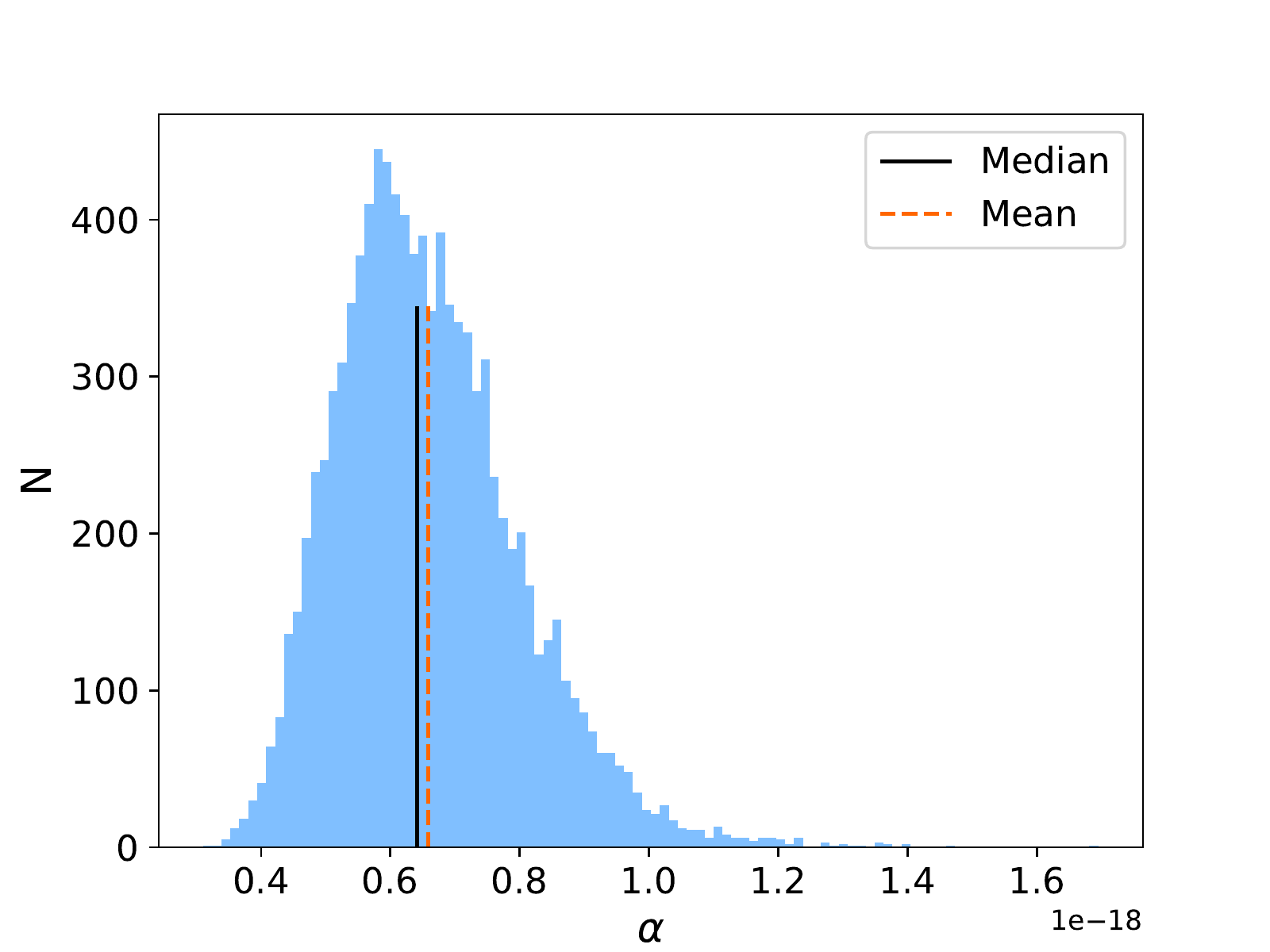}
    \caption{Distribution of obtained parameters from the blackbody fit to the 10,000 simulated data sets.}
    \label{fig:histparam}
\end{figure}

We have fit the blackbody model to each of the 10,000 data sets and we have obtained the two parameter distributions shown in Fig. \ref{fig:histparam}. The effective temperature distribution is a gaussian distribution, whose median and standard deviation is $\Teff$ = 910$\pm$50 K. The scaling factor is proportional to R$_{\rm bb} / $d, where R$_{\rm bb}$ is the radius of the blackbody and d the distance to the system. The mean value of the distribution of the scaling gives a rough estimate of R$_{\rm bb}\sim$ 10$^{11}$ cm $\sim$ 100 R$\oplus$.

In Fig. \ref{fig:disc_bbfit}, a random set of 500 blackbody fits is shown along with the IR excess data points, the photometry, the scaled white dwarf model and its blackbody fit and the sum of this fit and the disc model corresponding to the obtained parameters. We can see that the resolved photometry is reasonably well fitted with this model. The reduced $\chi^2$ of the fit to the IR excess is $\sim$2.5. 

\section{Discussion}
\label{sec:discussion}

The discovery of the disc around  WDJ1814$-$7354 was serendipitous, as the initial search in which this object was found was aiming to identify  white dwarf + ultracool dwarf pairs. This object was not found in previous searches for discs around white dwarfs. An explanation could be that the background objects made its IR excess unlikely to be due to a disc, until the photometry was recently deblended. This is an issue for \textit{WISE} selected white dwarfs with infrared excesses, as noted recently by \citet{Dennihy2020}.

\begin{figure}
    \includegraphics[width=\linewidth]{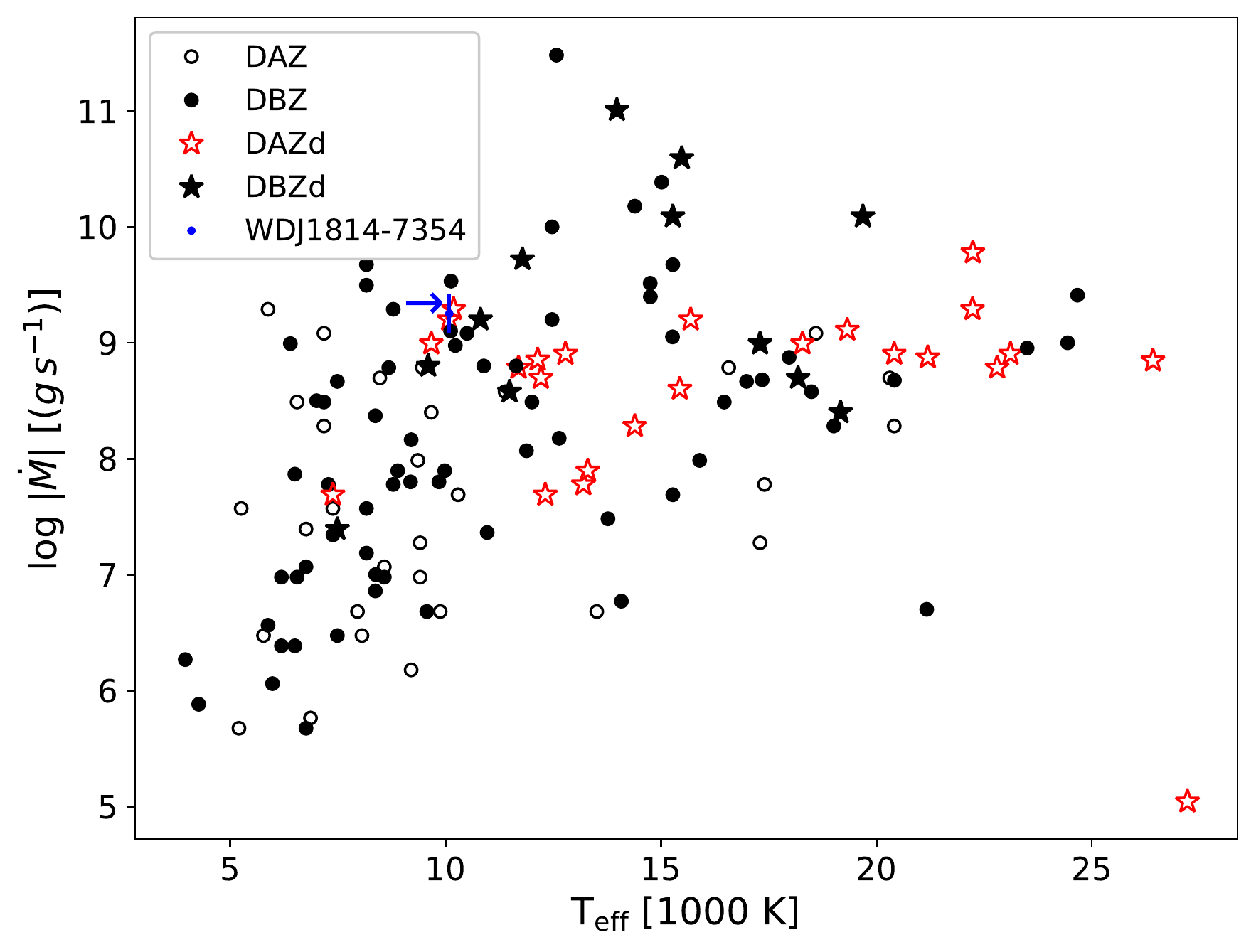}
    \caption{Time-averaged dust accretion rates vs. effective temperature of literature white dwarfs with the value for WDJ1814$-$7354 over-plotted with a blue dot symbol, and a blue arrow pointing at its location for visibility. DAZ- and DBZ-type stars are plotted as open and filled
circles, respectively, while objects with infrared excess are displayed as open and filled stars rather than circles. This figure is based on Fig. 10 from \citet{ReviewFarihi2016}, updated with data from \citet{Swan2019b} and \citet{Xu2019}.
}
    \label{fig:plotFarihi}
\end{figure}

The total estimated accretion flux of $\dot{\rm M} = (1.8\pm0.7)\times 10^{9}$\, g\,s$^{-1}$, presented in Table \ref{tab:abundances}, is higher than the average value for DAZ type objects with detected discs, (see table 6 of \citeauthor{Xu2019} \citeyear{Xu2019}), as can be also seen in Fig. \ref{fig:plotFarihi}. In this figure, we have placed WDJ1814$-$7354 in Fig. 10 of \citet{ReviewFarihi2016}, updated with data from \citet{Swan2019b} and \citet{Xu2019}, to compare it with other white dwarfs with measured accretion rates from the literature. It has the fourth highest accretion flux of the DAZd type (defined by \citeauthor{VonHippel2007} \citeyear{VonHippel2007} as white dwarf of DAZ type with a detected circumstellar disc), and this flux is higher than 83.8 per cent of all the other white dwarfs shown in the plot. We can see that WDJ1814$-$7354 is also one of the coolest (hence, with longer cooling ages) DAZ white dwarfs with discs known to date, with only 16 per cent of this sample having lower temperatures, and it is the third coolest of the known sample of 17 DAZd white dwarfs. 

\begin{figure}
    \includegraphics[width=\columnwidth]{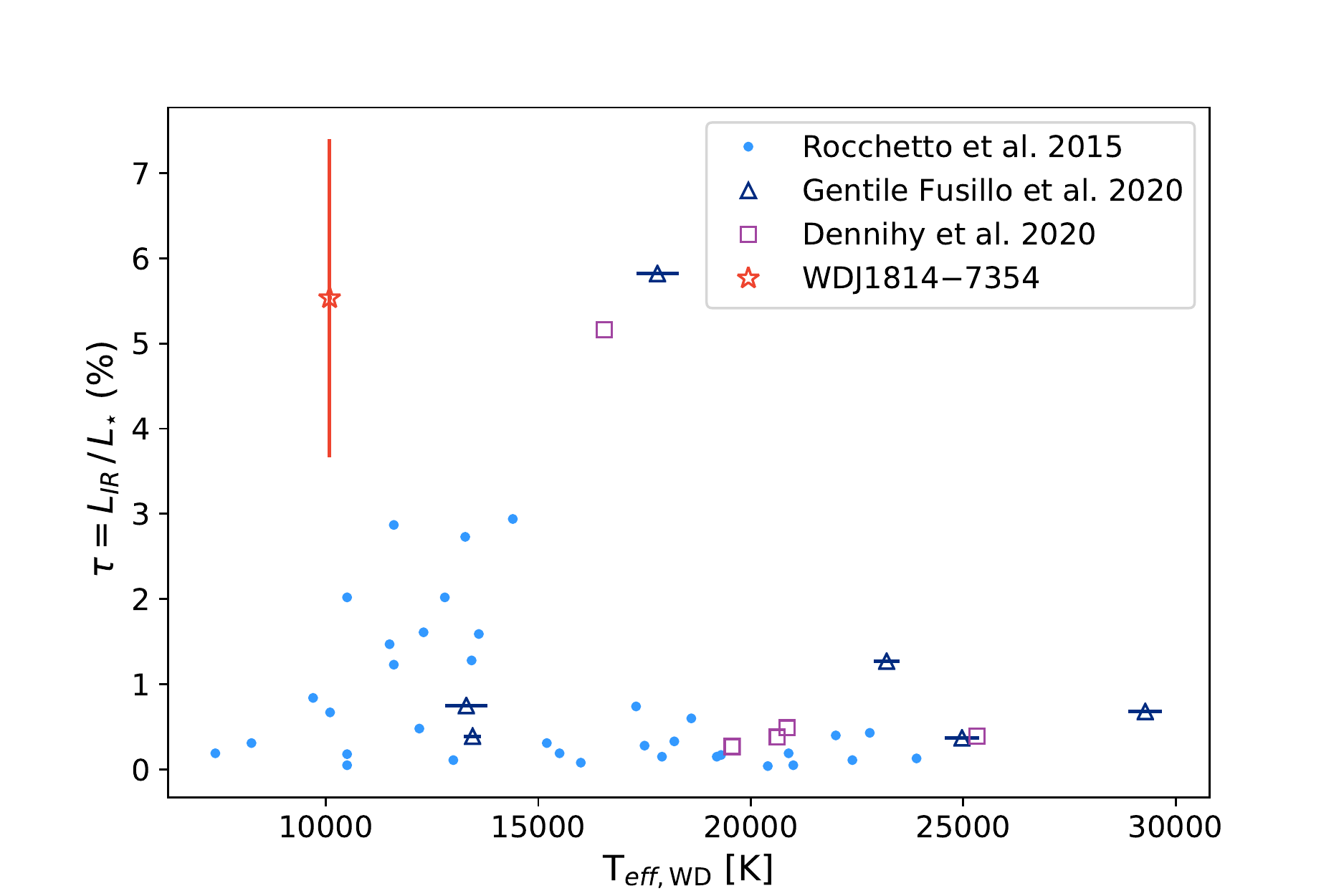}
    \caption{Percentages of fractional disc luminosity vs effective temperature of the central white dwarf for our system compared with data from \citet{Rocchetto2015}, \citet{GentileFusillo2020b} and \citet{Dennihy2020b}.
}
    \label{fig:fractionalluminosity}
\end{figure}

We have estimated the fractional disc luminosity ($\tau = L_{\rm IR}/L_*$) of our system by calculating the bolometric fluxes of the IR excess and the white dwarf. To do so, we have integrated the blackbody models of the IR excess and the white dwarf within the limits where the flux drops 99.99 per cent. The uncertainty for each bolometric flux has been estimated as the standard deviation of a distribution of 10,000 blackbody models created by randomly varying the model parameters within their errors. We have compared the fractional luminosity obtained with that of other systems from the literature in Fig. \ref{fig:fractionalluminosity}. We can see that WDJ1814$-$7354 disc luminosity fraction is among the largest of the discovered systems. In this plot it is the largest for cool white dwarfs (with $\Teff < 15,000$ K), although many systems from the literature do not have fractional luminosities available and were not included in this plot, and the available data for the rest does not include uncertainties, so the comparison has to be taken with caution. 

Finally, under the assumption that WDJ1814$-$7354 is currently accreting debris from a circumstellar reservoir in a steady state, i.e. with equilibrium between accretion and diffusion, we can estimate the accreted body composition, relative to Fe, via the following equation,  \citep[cf.][]{Koester2009, Gansicke2012}:

\begin{equation}
\frac{N(\mathrm{X})}{N(\mathrm{Fe})} = \frac{\dot{\rm M}(X) A(\mathrm{Fe})}{\dot{\rm M}(\mathrm{Fe})A(\mathrm{X})}
\label{eq:NXNFe}
\end{equation}
where $N(\mathrm{X})$ is the number abundance of element $\mathrm{X}$ and $A(\mathrm{X})$ its atomic weight. 

\begin{figure}
    \includegraphics[width=\columnwidth]{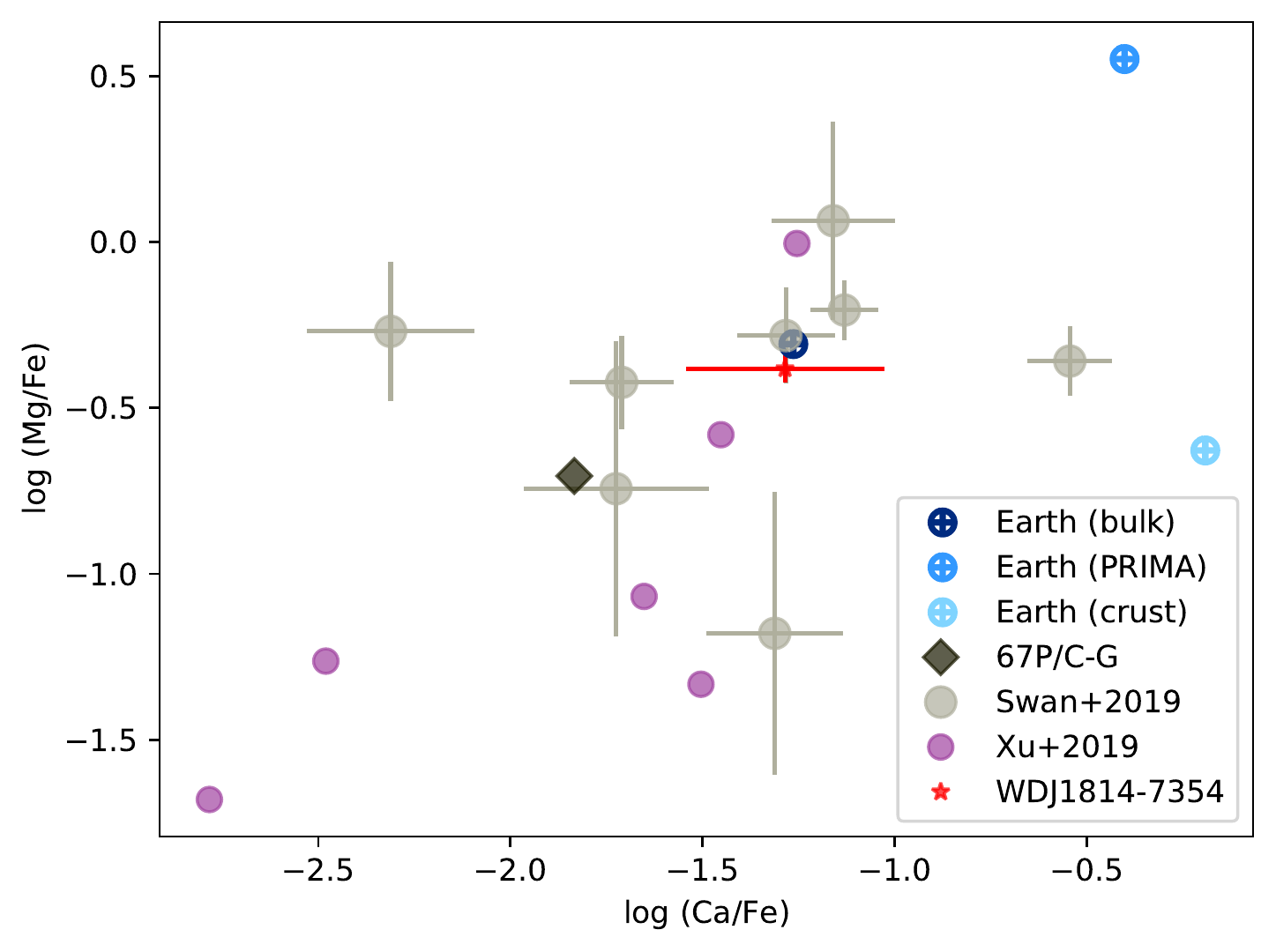}
    \caption{Comparison between the log mass abundance ratio of Mg and Ca for WDJ1814$-$7354 with white dwarfs and solar system objects, from \citet{Swan2019} and \citet{Xu2019}. The log mass abundance ratio is defined as log$_{10}N(\mathrm{X}) / N(\mathrm{Fe}$), obtained from Eq. \ref{eq:NXNFe}. Earth abundances are from the bulk silicate Earth (the primitive mantle, representing the composition of the upper layers of the Earth after the core separated, abbreviated PRIMA in the figure), bulk Earth and Earth crust.}
    \label{fig:abundances}
\end{figure}

Thus, we can compare the composition of the accreted material of WDJ1814$-$7354 with that of Solar system bodies and of other polluted white dwarfs from the literature (see Fig. \ref{fig:abundances}). A steady state can be assumed to be valid for white dwarf atmospheres that are characterised by short diffusion timescales, like for WDJ1814$-$7354, and systems that do not show line or disc variability. \citet{Veras2020, Girven2012, Farihi2012b} show that disc lifetimes are expected to be much longer than such diffusion timescales. In our case, the disc could have been building up material for $\sim$10 years, and it is expected not to show a change in its composition if the accretion timescale is longer than the diffusion timescale.

 The material being accreted by WDJ1814$-$7354 has similar metal ratios of Ca/Fe and Mg/Fe like those of bulk Earth, as can be seen in Fig.~ \ref{fig:abundances}. The upper limits available for other elements do not allow a wider comparison with the chemical profile of known Solar system bodies, as it has been done for other stars in the literature. We cannot accurately determine the total mass of the disc \citep[See e.g][]{Dufour2010} but we can estimate the mass of the accreted material. By multiplying the order of magnitude of the estimated lifetime of white dwarf discs, $10^5$ yrs \citep{Girven2012}, with our measured total accretion rate of detected elements of $\sim10^9$\,g\,s$^{-1}$, given in Table \ref{tab:abundances}, we obtain $10^{19}$ kg of accreted material, which is in the region of masses of Solar system asteroids such as 13 Egeria \citep{Baer2011} or 48 Bamberga \citep{carry2012}. This finding supports the evidence that rocky minor planetary bodies with Earth-like composition form around Sun-like stars elsewhere.

\section{Summary}
\label{sec:conclusions}

We have presented a spectroscopic and photometric analysis of the white dwarf WDJ1814$-$7354, an object originally identified to have a strong infrared excess in the 2MASS and \textit{WISE} catalogues. We confirmed this IR excess to be intrinsic to the white dwarf, and likely corresponding to the emission of a dusty disc around the star. The finding of Ca, Fe and Mg absorption lines in the X-SHOOTER spectrum of the white dwarf is further evidence of accretion from a dusty disc.  

We have compared the circumstellar metal lines in two spectra taken 8 years apart, finding no significant changes in radial velocities or in equivalent widths. This is not surprising, as the diffusion timescales derived in Section \ref{sec:abundances} for Mg, Ca and Fe go from 10 to 22 years, and as there has not yet been found unambiguously variability in white dwarf absorption metallic lines. Due to the blended photometry for this object in most IR surveys, we cannot study the potential variability of WDJ1814$-$7354 in the IR without follow-up observations.

The ratio of the disc and white dwarf luminosities is among the highest from the literature for a relatively cool white dwarf. From the analysis of the composition of the accreted material and the estimated total accretion rate, we suggest the possibility that a minor body was tidally disrupted into forming the disc of debris material around WDJ1814$-$7354.

\section*{Acknowledgments}

We thank the anonymous referee for a quick and adequately thorough revision of this manuscript. EGE acknowledges Astronomers for Planet Earth\footnote{https://astronomersforplanet.earth/} and recent works that raise awareness about CO$_2$ emissions from our professional activities \citep[See e.g][]{CO2-1, CO2-2} and commits to not travel anywhere by aeroplane for the purpose of promoting this paper. EGE also thanks Andrew Swan and astronomy PhD students in the University of Hertfordshire for helpful discussions. We thank Dr. Siyi Xu and Dr. Amy Bonsor for kindly sharing their recent X-SHOOTER and \textit{Spitzer} observations of this white dwarf. We thank Beth Klein and Jay Farihi for a helpful discussion that improved this article. 

 EGE and WJC are supported by a University of Hertfordshire PhD studentship. Part of this work was possible thanks to the support of the Royal Astronomical Society via the RAS grant E. A. Milne Travelling Fellowship awarded to EGE.  RR has received funding from the postdoctoral fellowship programme Beatriu de Pin\'os, funded by the Secretary of Universities and Research (Government of Catalonia) and by the Horizon 2020 programme of research and innovation of the European Union under the Maria Sk\l{}odowska-Curie grant agreement No 801370. FM acknowledges support from the NASA Postdoctoral Program at the Jet Propulsion Laboratory, administered by Universities Space Research Association under a contract with NASA. FM also acknowledges support from grant \#80NSSC20K0452 under the NASA Astrophysics Data Analysis Program. JCB acknowledges support from FONDECYT postdoctorado 3180716. LKR would like to acknowledge funding from The Science and Technology Facilities Council and the Institute of Astronomy, Cambridge. 
 
 Based on observations collected at the European Southern Observatory under ESO programmes 087.C$-$0639(B) (PI Dr. Avril Day-Jones) and 0103.C-0431(B) (PI Dr. Siyi Xu, Rogers et al. in prep.), and observations by \textit{Spitzer} space telescope, DDT program ID 14220, with PI Dr. Siyi Xu (Lai et al. in prep.). This work has made use of data from the European Space Agency (ESA) mission {\it Gaia} (\url{https://www.cosmos.esa.int/gaia}), processed by the {\it Gaia} Data Processing and Analysis Consortium (DPAC, \url{https://www.cosmos.esa.int/web/gaia/dpac/consortium}). Funding for the DPAC has been provided by national institutions, in particular the institutions participating in the {\it Gaia} Multilateral Agreement. The VISTA Data Flow System pipeline processing and science archive are described in \citet{Irwin2004}, \citet{Hambly2008} and \citet{Cross2012}. We have used data from the 6th data release (VHSDR6).

This research made use of {\sc{numpy}} \citep{numpy}, {\sc{scipy}} \citep{scipy}, {\sc{astropy}}, \citep{astropy}, {\sc{mat-plotlib}}, \citep{matplotlib}, {\sc{pandas}} \citep{pandas} and SAOImage DS9 \citep{ds9}.

\section*{Data Availability}

All data underlying this article is publicly available from the relevant observatory archive (see section \ref{sec:observations} and acknowledgements) or it is available on https://doi.org/10.5281/zenodo.4268013


\bibliographystyle{mnras}
\bibliography{References.bib}



\bsp	
\label{lastpage}
\end{document}